\documentclass[a4paper,12pt,oneside]{amsart}
\usepackage{graphicx}
\usepackage{amssymb}
\usepackage{caption2}
\usepackage{paralist}

\newtheorem{theorem}{Theorem}[section]

\newtheorem{prop}[theorem]{Proposition}
\newtheorem{rem}[theorem]{Remark}   

\theoremstyle{definition}

\newtheorem{example}[theorem]{Example}

\theoremstyle{remark}

\numberwithin{equation}{section}

\def\osc{ {\mathrm{osc}}}
\def\sem{ {\mathrm{sc}}}

\def\D{ {\scriptscriptstyle\mathrm{D}}}

\def\spec{ {\mathrm{Spec}}}

\def\OmegaO{\Omega^0}
\def\KGUE{ K_{\scriptscriptstyle\mathrm{GUE}}}
\def\KGOE{ K_{\scriptscriptstyle\mathrm{GOE}}}
\def\KGSE{ K_{\scriptscriptstyle\mathrm{GSE}}}
\def\DeltaO{\Delta_{\Omega}  }
\def\H{ {\mathcal{H}}}
\def\Gg{ {\mathbf{G}}}
\def\TH{ T_{\scriptscriptstyle\mathrm{H}}}
\def\D{ {\mathcal{D}}}
\def\S{ {\mathcal{S}}}
\def\A{ {\mathcal{A}}}
\def\Rr{ {   \mathcal{R}_{\scriptscriptstyle\mathrm{r}}   }}
\def\Rpr{ {   \mathcal{R}_{\scriptscriptstyle\mathrm{pr}}   }}
\def\Ri{ {   \mathcal{R}_{\scriptscriptstyle\mathrm{c}}   }}

\def\R{ {\mathcal{R}}}
\def\trace{ {\mathrm{Tr}}}
\def\nb{\bar{n}}
\def\Im{ {\mathrm{Im}}}
\def\Re{ {\mathrm{Re}}}

\def\Ham{ {\widehat{H}}}
\begin{document}

\title{SPECTRAL STATISTICS OF ``CELLULAR''  BILLIARDS}

\author{Boris Gutkin}
\address{
Fachbereich Physik,
Universit{\"a}t Duisburg-Essen
}
\email{boris.gutkin@uni-duisburg-essen.de}



\begin{abstract}
For a bounded  domain $\Omega^0\subset \mathbb{R}^2$ whose boundary contains a number of flat pieces $\Gamma_i$, $i=1,\dots l$
we  consider a family of   non-symmetric billiards  $\Omega$  constructed by patching several copies of $\Omega^0$ along $\Gamma_i$'s.  It is demonstrated  that the length spectrum of the periodic orbits  in $\Omega$ is degenerate with the multiplicities determined by a  matrix group $G$. We study
  the energy spectrum of the corresponding quantum billiard problem in $\Omega$ and show that it
can be  split  in a number of uncorrelated subspectra corresponding to a set of irreducible representations  $\alpha$  of  $G$. Assuming that the classical dynamics in  $\Omega^0$  are  chaotic, we derive    a semiclassical trace formula for each spectral component  and show that  their  energy level statistics are the same as in standard Random Matrix ensembles. Depending on whether ${\alpha}$ is real, pseudo-real or complex, the spectrum   has either    Gaussian Orthogonal, Gaussian Symplectic or Gaussian Unitary  types of statistics, respectively.

\end{abstract}

\maketitle

\section{Introduction}

According to the Bohigas-Giannoni-Schmit conjecture \cite{bgs} the  energy spectrum  of a generic  Hamiltonian  system with  classically  chaotic dynamics is distributed in the same way as spectra  of the standard   random matrix  ensembles within  the same symmetry class. In particular, the spectral statistics of   spinless single-particle  systems with time-reversal invariant  classical chaotic dynamics can  usually be described by the Gaussian Orthogonal Ensemble (GOE), or by the Gaussian Unitary Ensemble (GUE), if the time-reversal invariance is broken. For chaotic single-particle  systems with half-integer spin the spectral statistics are typically the same as in  the Gaussian Symplectic Ensemble (GSE).
There exist, however, a few notable exceptions from this rule. It is known, for instance, that  the  presence of additional  symmetries might lead to a change of the spectral statistics \cite{br1, lss}. If the system possesses
a discreate symmetry group $H$, its spectrum   can be  split into a number of uncorrelated subspectra, where each spectral component  corresponds to an   irreducible representation $\alpha$ of $H$ \cite{rob,lau}.
The distribution of the energy levels $\{ E^{(\alpha)}_n\}$ within each sector depends then on the type of the representation $\alpha$ \cite{kr,ms}.
   For real  and pseudo-real representations  the corresponding spectral statistics are
of GOE and of GSE type, respectively.\footnote{I am indebted to C. Joyner, S.\ M\"uller and
 M. Sieber
 for pointing out to me that pseudo-real representations of the symmetry group give rise to GSE type of spectral statistics.
} If, on the other hand,   $\alpha$  is complex, then   
the corresponding spectral statistics are      of GUE type. As a result,  even time-reversal invariant
   systems  might contain subspectra of GUE type provided  $H$ has complex irreducible representations \cite{ lss, kr}. 
Other examples of non-standard  spectral statistics are provided by the Laplacian eigenvalues of certain  arithmetic surfaces of constant negative curvature \cite{bog1, bog2}, as well as  by some linear hyperbolic automorphisms of the
2-torus (cat maps) \cite{cat1, cat2}.  Anomalous spectral statistics   also appear  in some chaotic systems with several ergodic components \cite{robnik, moj}.

\begin{figure}[htb]
\begin{center}{
(a)
\includegraphics[height=3.8cm]{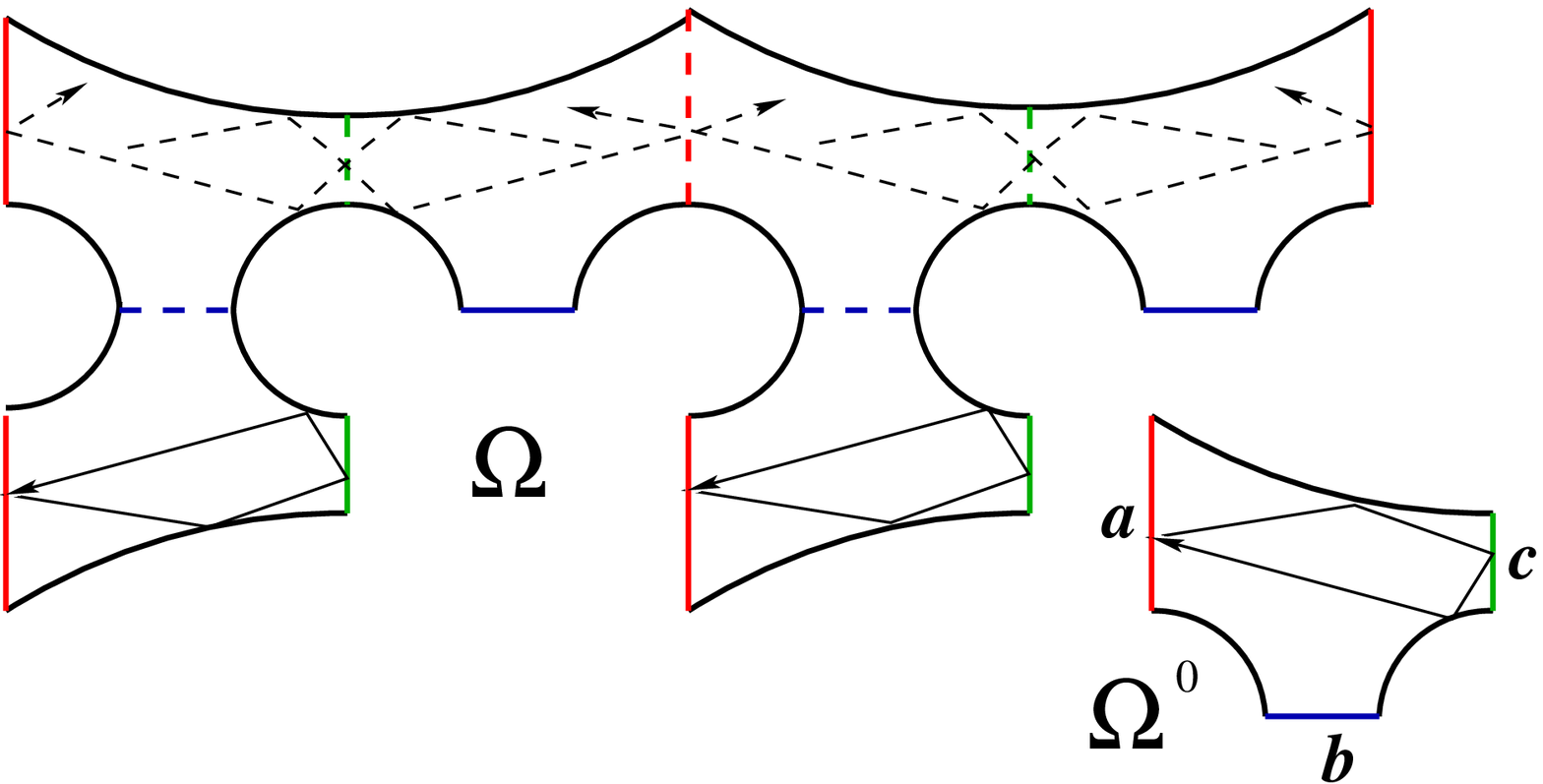}
\hskip 0.1cm
(b)
\includegraphics[height=3.0cm]{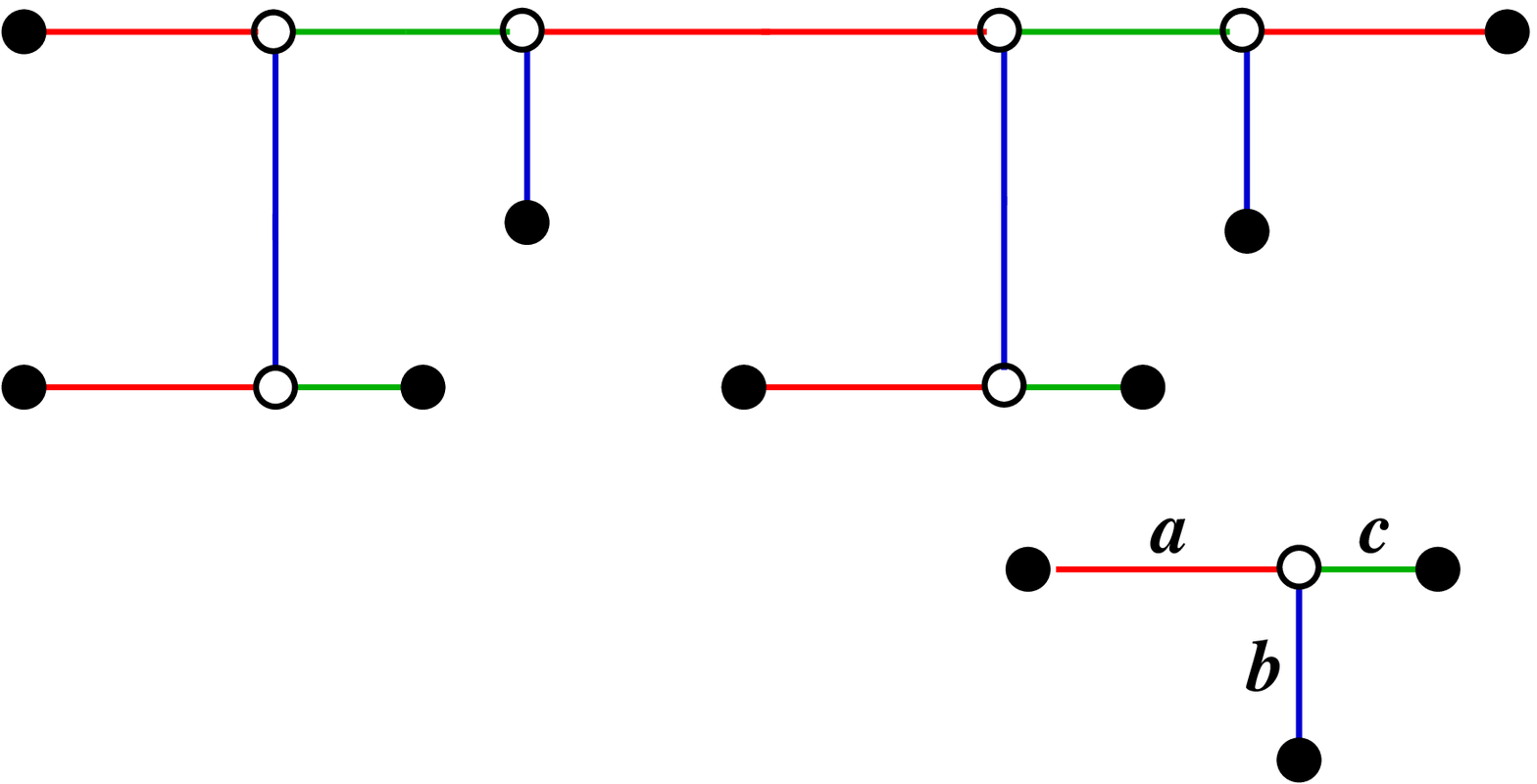}
}\end{center}
\caption{ \small{Construction of  cellular (quantum) billiards (a) and  (quantum) graphs (b). The billiard $\Omega$ is obtained by connecting  six copies of $\OmegaO$  along its flat sides (coloured lines). At each flat part of the billiard boundary   $\partial\Omega$ we impose either  Dirichlet or Neumann boundary conditions. } }\label{fig1}
\end{figure}

 From  the semiclassical point of view   deviations from the spectral universality   can be  always traced   to a certain 
anomaly   in the length spectrum of the classical {\it periodic orbits} (PO). For instance,  additional geometrical symmetries  imply degeneracies     in the  length spectrum  of periodic orbits. Note, however, that even in the absence  of  geometrical symmetries  degeneracies in the  length spectrum might  exist.   This    happens, for instance, in the case of arithmetic surfaces of negative curvature, where   large multiplicities of the periodic orbits lead to the spectral statistics reminiscent of   the Poissonian  distribution. In the present work we consider another     class of 
non-symmetric billiards  whose  length spectrum of periodic orbits is  degenerate. These billiards are constructed  in the following way. 
 Let $\OmegaO$ be a bounded  domain on 
$\mathbb{R}^2$ with the boundary  $\partial \OmegaO$
  containing $l$ flat pieces  (i.e., line segments) $\Gamma_{i}$, $i=1,\dots l$. The billiard  domain $\Omega$  is  then builded up by taking $N$  copies of $\OmegaO$ and connecting them with the opposite orientations along  $\Gamma_{i}$'s, see fig.~\ref{fig1}a. The resulting domain $\Omega$ is equipped with the flat metric and in some  cases with an appropriate choice of $\OmegaO$ can be  embedded into $\mathbb{R}^2$.\footnote{Note, however, that an embedding  of $\Omega$ in $\mathbb{R}^2$  is not always possible and we actually do not require it.} In what follows we will refer to  any such  domain $\Omega$  as {\it cellular billiard}     and to  $\OmegaO$ as   {\it fundamental cell}.  It should be noted that the same construction can be carried out in any dimension.  In particular, starting from a one-dimensional  fundamental cell one can also construct (quantum) {\it cellular  graphs}, see fig.~\ref{fig1}b. 
It is easy to see that the classical dynamics in  $\Omega$  is intimately connected with the classical dynamics in $\OmegaO$.  Specifically, for every periodic orbit in  $\Omega$ there exists a corresponding  periodic orbit in  $\OmegaO$ of the same length. The opposite, however, is not always true: a periodic orbit in  $\OmegaO$, in general,  gives  rise to a number (which can be also zero) of periodic orbits in $\Omega$. As a result, $\OmegaO$ and $\Omega$ posses the same  length spectrum of periodic orbits but with different multiplicities, see fig.~\ref{fig1}a.

Consider now  the corresponding quantum billiard problem in   $\Omega$:
\begin{equation}
 -\Delta_{\Omega}\varphi=\lambda\varphi,
\end{equation}
where the function  $\varphi$ satisfies some boundary conditions at  $\partial\Omega$ and $\Delta_{\Omega}$ stands for the corresponding Laplacian. Note by passing, that  the  spectral problem for such billiards  has previously attracted  an attention in connection to the famous question of M. Katz \cite{kac}: ``Can one hear the shape of a drum?''. It was  shown  \cite{gordon} that starting from the same initial domain $\OmegaO$ one can construct  in certain cases a pair of non-isometric  domains  $\Omega$,    $\Omega'$  such that the spectra of  $\Delta_{\Omega}$ and $\Delta_{\Omega'}$ coincide (for a similar construction of isospectral graphs, see \cite{mojuzy,rami}).  Here we are rather interested  in the  spectral properties of $\Delta_{\Omega}$ for a general cellular billiard $\Omega$. Note that, generically,  $\Omega$ does not have  geometric symmetries. On the other hand,  the length spectrum of periodic orbits in $\Omega$ is degenerate and  one might  suspect that   the   energy levels statistics of $\Delta_{\Omega}$ exhibit an  anomaly. 

 The main goal of this paper is to give a precise description of the spectral structure of $\Delta_{\Omega}$, based on the assumption of chaotic dynamics in the fundamental cell $\OmegaO$. As we will show,  the situation here is reminiscent  of that encountered in systems with   geometrical  symmetries. Namely,  $\Delta_{\Omega}$ can be split in a number of subspectra: 
\begin{equation}
 \Delta_{\Omega}=\bigoplus_{\alpha\in\R} \Delta^{(\alpha)}_{\Omega},
\end{equation}
where the sum runs over a subset  $\R$ of   irreducible representations for certain matrix group $G$. Note that in  general $G$ is not a symmetry group of the billiard domain, but rather a structure group which determines  multiplicities of periodic orbits in $\Omega$. (The exact definition of $G$ and  the relevant  set  $\R$ of its irreducible representations will be provided in the body of the paper.)  As  in the case of geometrical symmetries, the spectral statistics of each sector $\Delta^{(\alpha)}_{\Omega}$ turns out to be determined by the type of the   representation $\alpha$ -- the statistics are of GUE type, if the representation is complex and of GOE,
 GSE  types,
if the representation is real or pseudo-real, respectively.  Furthermore, we show that  the spectra of different   sectors $\alpha$ are  uncorrelated.

The paper is organised as follows. In Sec.~2 we consider the semiclassical trace formula for $\Delta_{\Omega}$ and express the  multiplicities $\eta_\gamma$  of the periodic orbits $\gamma$   in $\Omega$ through the traces of certain class of permutation matrices. We then show that  $\eta_\gamma$ combined with the  phases (resulting from the Dirichlet boundary conditions) can be represented as characters of the standard representation for some matrix  group $G$.  In Sec.~3 we apply the  trace formula to obtain  spectral correlations of   $\Delta_{\Omega}$. We show that  the resulting spectral statistics are, in general,   mixtures of GUE, GOE
  and GSE
types of distributions. In Sec.~4 we demonstrate that the spectrum of  $\Delta_{\Omega}$ can be split into a number of subspectra corresponding to  irreducible representations of $G$ and derive a semiclassical trace formula for  these subspectra in Sec.~5.  Finally, the conclusion is presented in Sec.~6.

\section{Semiclassical trace formula}
Let $\OmegaO$  be a bounded domain on   $\mathbb{R}^2$, with a piecewise smooth boundary $\partial\OmegaO$ containing $l$ flat pieces $\Gamma_i$, $i=1,\dots l$. We will consider the associated quantum Hamiltonian $\Ham^{(0)}:=-(\hbar^2/2)\Delta_{\OmegaO}$, where $\hbar$ is Planck's constant and $\Delta_{\OmegaO}$ is the Laplacian in $\OmegaO$ with the Neumann boundary conditions at $\partial\OmegaO$. The spectrum $\{ E^{(0)}_j\}^{\infty}_{j=0}$,  $E^{(0)}_j=\hbar^2\lambda^{(0)}_j/{2}$ of $\Ham^{(0)}$
is then defined by the solutions of the following eigenvalue problem
\begin{eqnarray} -\Delta\varphi^{(0)}_j = \lambda^{(0)}_j\varphi^{(0)}_j,  \qquad \varphi^{(0)}_j \in L^2(\OmegaO),  \quad \partial_n\varphi^{(0)}_j|_{{\partial\OmegaO}}=0.\label{qbilliard0}
\end{eqnarray}
Now, take  $\OmegaO$ as the fundamental cell and  construct a cellular  billiard  $\Omega$   by means of the procedure described in the previous section. Note that, in general, the billiard boundary $\partial\Omega$     contains a number of flat pieces corresponding to unpaired sides $\Gamma_i$ of the copies of $\OmegaO$, see fig.~\ref{fig1}.
We will study  the  quantum billiard problem in   $\Omega$  for mixed Dirichlet-Neumann boundary conditions at  flat pieces  of $\partial\Omega$. For the sake of concreteness we fix   the Neumann   boundary conditions at the rest  of the boundary. Let $\partial\Omega_i$, $i=1,\dots \ell$ be the flat pieces of $\partial\Omega$ with  the Dirichlet   boundary conditions and let $\overline{\partial\Omega}=\partial\Omega\backslash\cup_{i}^{\ell}\partial\Omega_i$ be the remaining part of the boundary, then the eigenvalue problem 
\begin{eqnarray} -\Delta\varphi_j = \lambda_j\varphi_j,&& \,\,\, \varphi_j \in L^2(\Omega), \nonumber \\
   \varphi_j|_{\partial\Omega_k}=0, \,\, k=1,\dots \ell,&& \,\,\,\partial_n\varphi_j|_{\overline{\partial\Omega}}=0, \label{qbilliard}
\end{eqnarray}
defines the  Laplace operator $\Delta_{\Omega}$ and the energy levels $\{ E_j\}^{\infty}_{j=0}$, $E_j=\hbar^2\lambda_j/2$ of the corresponding quantum Hamiltonian $\Ham:=-(\hbar^2/2)\Delta_{\Omega}$. 
 In what follows we will consider the   spectral density functions for the quantum billiards in $\OmegaO$,  $\Omega$: 
\begin{equation}
 d^{(0)}(E)= \sum_{n=0}^{\infty}\delta(E-E^{(0)}_n), \qquad  d(E)= \sum_{n=0}^{\infty}\delta(E-E_n) \label{spectraldensity}
\end{equation}
under the assumption that the classical dynamics in the fundamental cell $\OmegaO$ are chaotic. 

The spectral function $d^{(0)}(E)=\bar{d}^{(0)}(E)+d^{(0)}_{\mathrm{osc}}(E)$ can be split into the  smooth $\bar{d}^{(0)}(E)$  and the oscillating part whose semiclassical form is  given by the Gutzwiller trace  formula \cite{haake}:
\begin{equation} d^{(0)}_{\mathrm{osc}}(E)= \frac{1}{\pi\hbar}\Re\sum_{\gamma\in \mathrm{PPO(\OmegaO)}}   \A_\gamma\exp{\left(\frac{i}{\hbar} S_\gamma(E)\right)} + \Bigl\{\substack{\text{Contribution from}\\ \text{repetition of p.o.}}\Bigr\}. 
\label{traceformulaO}\end{equation}
Here   the sum  runs over the set of all {\it prime periodic orbits} (PPO) in $\OmegaO$ and $S_\gamma$, $\A_\gamma$  are   the  action (including Maslov indices) and the  stability factor of $\gamma$.
In eq.~(\ref{traceformulaO}) we singled out the  contribution of the prime periodic orbits, as only these orbits are relevant for the spectral correlations. The contributions  from the periodic orbits with  a number of repetitions turn out to be suppressed by their large instability factors, see e.g.,  \cite{haake}.

Analogously, one can express the spectral density of states for the billiard $\Omega$ in terms of its periodic orbits.
 Since each periodic orbit of $\Omega$ is also a periodic orbit of $\OmegaO$, the oscillating part of $d(E)=\bar{d}(E)+d_{\mathrm{osc}}(E)$ can be represented as a sum over the periodic orbits $\gamma$ of the billiard $\OmegaO$:   
\begin{equation} d_{\mathrm{osc}}(E)= \frac{1}{\pi\hbar}\Re\sum_{\gamma\in \mathrm{PPO(\OmegaO)}}  \chi_\gamma \A_\gamma\exp{\left(\frac{i}{\hbar} S_\gamma(E)\right)}+ \Bigl\{\substack{\text{Contribution from}\\ \text{repetition of p.o.}}\Bigr\},  \label{traceformula}\end{equation}
where    $\chi_\gamma=(-1)^{k_\gamma} \eta_\gamma$ with $k_\gamma$ being the number of times $\gamma$ hits the pieces of the boundary with the Dirichlet boundary conditions and $\eta_\gamma\in \mathbb{Z}^+$ being the multiplicity of $\gamma$.  
Note that the forms of (\ref{traceformulaO}) and (\ref{traceformula}) are almost identical   with a noticeable difference  of additional multiplicity factors $\chi_\gamma$ in the last expression. Let us show now that these factors can be identified as characters  of the standard representation for some matrix  group.

Let $\{ \Gamma_k,  k=  1, \dots l \}$ be the set  of flat components at the boundary of the domain $\OmegaO$. For each such component $\Gamma_k$ we define an associated $N\times N$ matrix $\sigma^{(k)}$ in the following way. 
Let  $\OmegaO_{i}$, $i=1,\dots N$, be $N$ copies of    $\OmegaO$ which compose the billiard $\Omega$.
Then $\sigma^{(k)}_{i,j}=1$ for $i\neq j$ if  $\OmegaO_i$ is  connected to  $\OmegaO_j$ through the side $\Gamma_k$, 
$\sigma^{(k)}_{i,i}=1$ (resp. $\sigma^{(k)}_{i,i}=-1$) if  the boundary component $\Gamma_k$ of $\partial\OmegaO_i$  belongs to the boundary $\partial\Omega$ with the Neumann boundary conditions (resp. the Dirichlet boundary conditions)  and $\sigma^{(k)}_{i,j} =0$, otherwise. The set of matrices  $\{ \sigma^{(k)},  k=  1, \dots l \}$ generates then the group $G$ with the multiplication operation given by the standart matrix product. In particular, for purely Neumann boundary conditions on  $\partial\Omega$, $\sigma^{(k)}$ are just permutation matrices and $G$ is isomorphic to a subgroup of the permutation group of  $N$ elements. 
Since $G$ is a matrix group, it admits the standard representation $\rho$, such that $\rho_{i,j}(\sigma)=\sigma_{i,j}$ for any  $\sigma\in G$. It is  straightforward to see that   the multiplicity factors $\chi_\gamma$ can be expressed through the charactors of $\rho$. Given a periodic orbit $\gamma$ in  $\OmegaO$, denote $\Gamma_{k_1}\Gamma_{k_2}\dots \Gamma_{k_{\bar{n}}}$, $k_i\in \{ 1, \dots l \}$ the (time) ordered  sequence of flat pieces of $\partial \OmegaO$ in which the billiard ball flying along $\gamma$ hits the boundary, then 
 \begin{equation}
 \chi_\gamma=\trace{\sigma_\gamma},\qquad \sigma_\gamma=\prod_{i=1}^{\nb}\sigma^{(k_i)}\in G. \label{sigmadef}
\end{equation}
The above formula can be  interpreted in the following way. Any periodic trajectory $\gamma$ in $\OmegaO$  passing through a    point $x\in \OmegaO$ gives rise to $N$ trajectories  $\{\gamma_1,\dots \gamma_N\}$ in $\Omega$. Each trajectory $\gamma_j$  starts at the   point $x_j$, where $x_i\in\OmegaO_{i}$, $i=1\dots N$ are the lifts of  $x$ on $\Omega$. Since $\gamma$ starts and ends at the same point, the 
 set of  endpoints of $\gamma_1,\dots \gamma_N$ 
is, in fact, a permutation  of the initial points $\{x_1,\dots x_N\}$. Specifically,  $x_j$ is the end point of  $\gamma_i$ if $(i,j)$'s element of $\sigma_{\gamma }$ is nonzero.  As a result, by setting all nonvanishing elements of $\sigma_\gamma$ to $+1$   we obtain a  permutation  matrix of initial and final conditions, where  the number of  units at the diagonal  defines the number of periodic  trajectories among $\{\gamma_1,\dots \gamma_N\}$. Furtheremore,  having the elements of $\sigma^{(k)}$ with both negative and positive signs allows us to take in account different boundary conditions on $\partial \Omega$ which are relevant for the semiclassical trace formula.

\begin{figure}[htb]
\begin{center}{
(a)
\includegraphics[height=3.8cm]{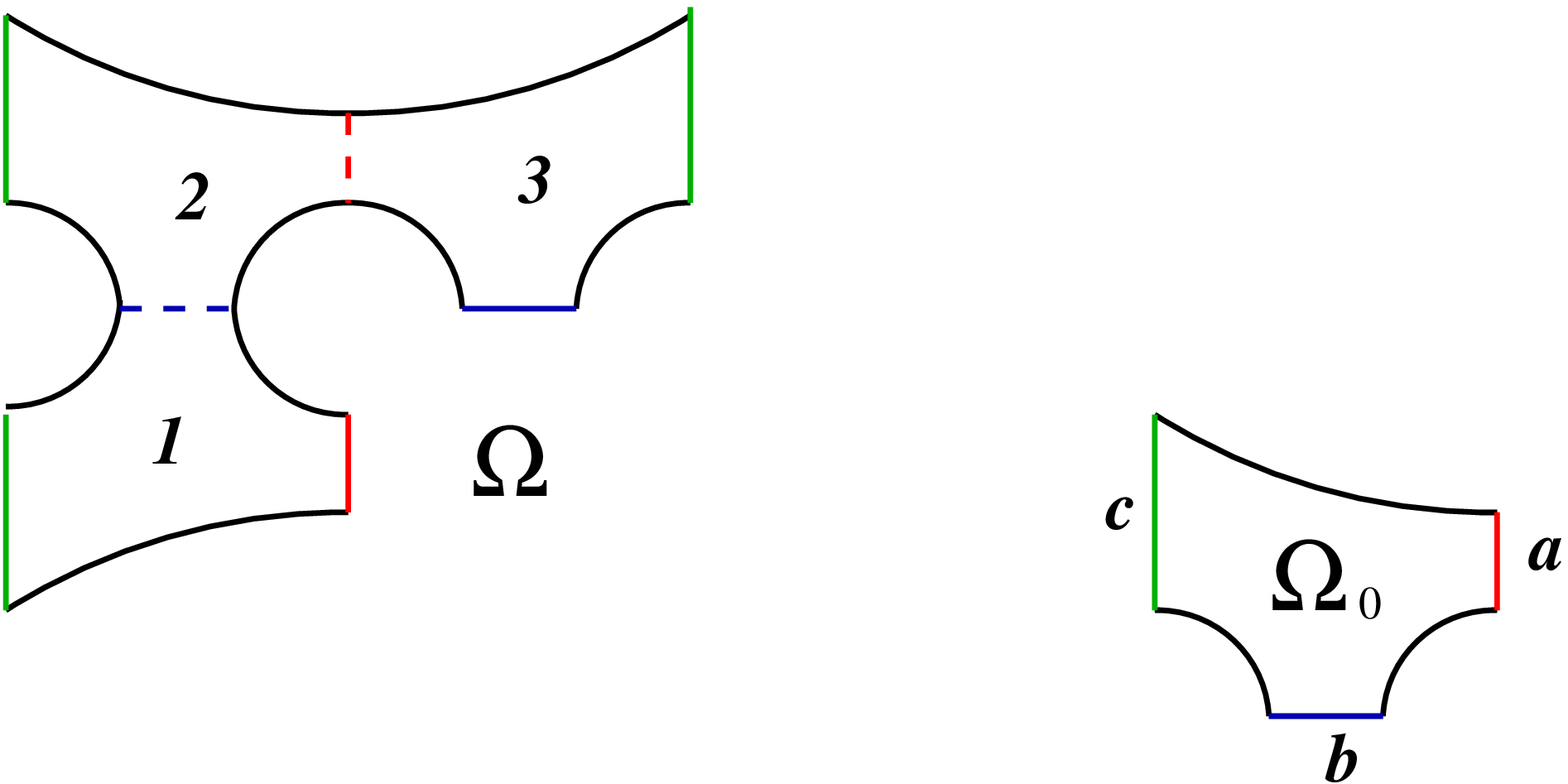}
\hskip 1.0cm
(b)
\includegraphics[height=3.3cm]{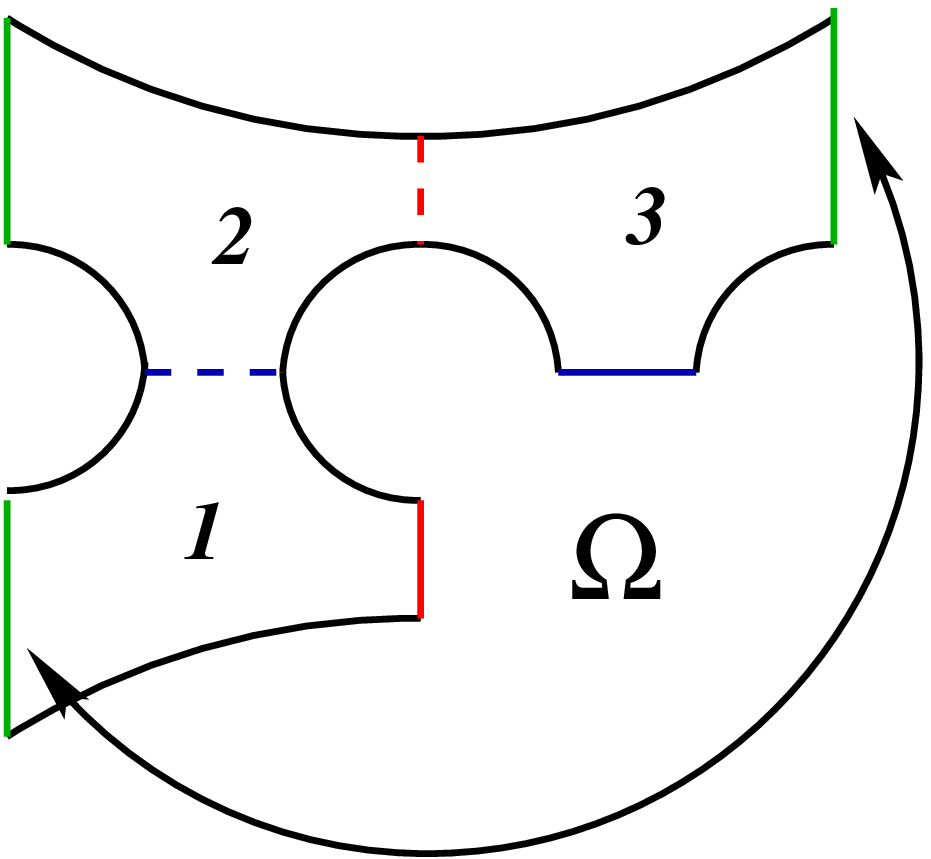}

}\end{center}
\caption{ \small{Pictures of two billiards considered in Example 2.1. Note that in the second case the first and the third copy of $\Omega_0$ are connected along the flat pieces of the boundary (marked by green colour), as shown by the arrow. }  }\label{fig}
\end{figure}

As an example, let us  consider two  billiards shown in  fig.~\ref{fig} with the Dirichlet boundary conditions on the flat parts of the boundary $\partial\Omega$. 
\begin{example}
 For the billiard  in fig.~\ref{fig}a the generating matrices are given by
\[\sigma_a=\begin{pmatrix}-1&0&0\\ 0&0&1\\0&1&0\end{pmatrix}, \quad \sigma_b=\begin{pmatrix}0&1&0\\ 1&0&0\\0&0&-1\end{pmatrix}, \quad \sigma_c=\begin{pmatrix}-1&0&0\\ 0&-1&0\\0&0&-1\end{pmatrix}.\]
The group  generated by  $\sigma_a,\sigma_b,\sigma_c$  contains $10$ elements:  $G=\{\sigma_0 \sigma,\sigma_1 \sigma|  \sigma\in G_0\}$, where 
\[ \sigma_0=-\sigma_1= \begin{pmatrix}1&0&0\\ 0&1&0\\0&0&1\end{pmatrix} \quad \mbox{ and } G_0=\{\sigma_0,  \sigma_a,  \sigma_b,  \sigma_a\sigma_b, \sigma_b\sigma_a\}.\]
Note also that $G_0$ is isomorphic to the group of permutations of three elements.

For  the billiard shown in  fig.~\ref{fig}b there is an additional connection between first and third copies of the fundamental cell and the generators of the group $G$ are given by: 
\[\sigma_a=\begin{pmatrix}-1&0&0\\ 0&0&1\\0&1&0\end{pmatrix}, \quad \sigma_b=\begin{pmatrix}0&1&0\\ 1&0&0\\0&0&-1\end{pmatrix}, \quad \sigma_c=\begin{pmatrix}0&0&1\\ 0&-1&0\\1&0&0\end{pmatrix},\]
and   $G=\{\sigma_i \sigma| i=0,1,2,3; \sigma\in G_0\}$, where $G_0$ is defined as above, with
\[ \sigma_0= \begin{pmatrix}1&0&0\\ 0&1&0\\0&0&1\end{pmatrix}, \sigma_1= \begin{pmatrix}1&0&0\\ 0&-1&0\\0&0&-1\end{pmatrix}, \sigma_2= \begin{pmatrix}-1&0&0\\ 0&-1&0\\0&0&1\end{pmatrix}, \sigma_3= \begin{pmatrix}-1&0&0\\ 0&1&0\\0&0&-1\end{pmatrix} .\]
\end{example}

\section{Spectral correlations}

We  now proceed with the calculations of the form factor 
\[K(\tau)= \int_{-\infty}^{+\infty} R(x/\bar{d})e^{-2\pi i\tau x}dx\]
 for the two-point correlation function  of the spectral level density: 
\begin{equation}
  R(\varepsilon)=\frac{1}{\bar{d}^2}\left\langle d\left(E+\frac{\varepsilon}{2}\right) d\left(E-\frac{\varepsilon}{2}\right) \right\rangle_E -1.
\end{equation}
Using the semiclassical expression (\ref{traceformula}) for the density of states and expanding the actions $S (E\pm\frac{\varepsilon}{2})\approx S(E)\pm\frac{\varepsilon}{2}T(E)$ up to the linear term one  obtains 
  \begin{equation}
 K(\tau)= \frac{1}{\TH^2}\left\langle\sum_{\gamma,\gamma'} \chi_{\gamma}\chi_{\gamma'}  \A_{\gamma} \A^*_{\gamma'} \exp{\left(\frac{i}{\hbar} (S_{\gamma}-S_{\gamma'})\right)}\delta\left(\tau-\frac{(T_{\gamma}+T_{\gamma'})}{2\TH}\right)\right\rangle_E,  \label{semcorrelator}
\end{equation}
where  $T_\gamma, T_{\gamma'}$ stand for  periods of  $\gamma, \gamma'\in\mathrm{PPO(\OmegaO)}$,  and   $\TH=2\pi\hbar\bar{d}$ is the Heisenberg time for $\Omega$. Note that the spectral form factor $K_0(\tau)$ of the quantum billiard in $\OmegaO$  can be expressed  in a similar way  by setting  all multiplicity factors $\chi_{\gamma},\chi_{\gamma'}$  in 
eq.~(\ref{semcorrelator}) to one and rescaling $\TH$ by the factor $1/N$: 
\begin{equation}
 K_0(\tau)=\frac{N^2}{\TH^2}\left\langle\sum_{\gamma,\gamma'}\A_{\gamma} \A^*_{\gamma'} \exp{\left(\frac{i}{\hbar} (S_{\gamma}-S_{\gamma'})\right)}\delta\left(\tau-\frac{(T_{\gamma}+T_{\gamma'})}{2\TH/N}\right)\right\rangle_E.\label{semcorrelatorO}
\end{equation}
 In what follows we are going to establish connection between $K_0(\tau)$ and 
$K(\tau)$.

The semiclassical expressions (\ref{semcorrelator},\ref{semcorrelatorO}) can be  used in order to calculate $K(\tau)$, $K_0(\tau)$ perturbatively, as  functions of the parameter  $\tau$. The leading order $\tau^1$ contribution   can be  obtained using  so-called diagonal 
approximation, where only pairs of the same  periodic  orbits   are considered \cite{berry}. The next order $\tau^2$ term is due to  the contribution of pairs of periodic orbits with one selfencounter (Sieber-Richter pairs) \cite{rs}.
In the same spirit   the higher order terms $\tau^{n+1}$ can be obtained  from   the correlations of non-identical trajectories with $n$   selfencounters \cite{haake1}.  Such correlating orbits can be organised into families according to their  topological structure.  Each family then, includes periodic orbits with  close actions  which systematically contribute into the sum. As a result, the form factor  $K_0(\tau)$  can be written in the perturbative form as:
\begin{equation}
 K_0(\tau)=\sum^{\infty}_{n=1}\tau^n c_n, \qquad c_n= \sum_{s\in \S_n}k_n^{(s)},  \label{structure}
\end{equation}
where  the last sum runs over the set $\S_n$ of topologically different structures of periodic orbits having  $n$ encounters. For  generic form  of the length spectrum of periodic orbits in chaotic systems with time reversal invariance, the contribution $k_n^{(s)}$ for each structure $s\in\S_n$   has been explicitly calculated in  \cite{haake1}  and   shown to reproduce  RMT result:

\[ \mathbf{GOE:} \,\, c_1=2, \,\, c_{n+1}=\frac{(-2)^n}{n}, \,\,n\geq 1; \quad \mathbf{GUE:}  \,\,c_1=1, \,\, c_{n+1}=0,  \,\, n\geq 1;\]
\[ \mathbf{GSE:} \,\, c_1=\frac{1}{2}, \,\, c_{n+1}=\frac{1}{4n}, \,\,n\geq 1.\]

To calculate the spectral form factor $K(\tau)$ for the billiard  $\Omega$ we will assume  that for long periodic trajectories  the multiplicity factors $\chi_{\gamma}$ 
 do not  correlate with the actions $S_\gamma$.     It follows then  by  (\ref{semcorrelator}) and (\ref{semcorrelatorO}) that  
      
\begin{equation}
 K(\tau)=\frac{1}{N}\sum^{\infty}_{n=1}(N\tau)^n c'_n, \qquad  c'_n=\sum_{s\in \S_n}\D_n^{(s)}k_n^{(s)}, \label{formfac1}
\end{equation}
where $k_n^{(s)}$ are  as in eq.~(\ref{structure}), and
\[\D_n^{(s)}=\left\langle \chi_{\gamma}\chi_{\gamma'} \right\rangle_{s},\]
 with the average   over all periodic orbits  $\gamma,\gamma'$ having the same topological structure $s\in\S_n$ of encounters.
Since $\chi_{\gamma}$,  $\chi_{\gamma'}$ are characters of  two group elements $\sigma_{\gamma}, \sigma_{\gamma'}\in G$ the above average over periodic trajectories can be  substituted with the average over the set $G^{(s)}=\{(\sigma, \bar\sigma)\}$ of pairs $\sigma, \bar\sigma\in G$ compatible with the  structure  $s$ of correlating periodic orbits:
 \begin{equation}  
\D_n^{(s)}=\frac{1}{|G^{(s)}|}\sum_{\sigma, \sigma'\in G^{(s)}}\chi(\sigma)\chi(\sigma'),\label{dnformula}
\end{equation}
where the normalisation factor  $|G^{(s)}|$ is the number of pairs in $G^{(s)}$.

 It follows  from eq.~(\ref{formfac1}) that, in order to evaluate  $K(\tau)$ one only need to know    the coefficients $\D_n^{(s)}$.  Below we show how to calculate $\D_n^{(s)}$   for a given structure $s$ of the  correlating periodic orbits.

\subsection{Diagonal approximation}

For the diagonal approximation  the two trajectories $\gamma, \gamma'$ coincide and we have: $G^{(s)}=\{(g,g)|\,  g\in G\}$. This yields
\begin{equation}
 \D_1=\frac{1}{|G|} \sum_{\sigma\in G} (\chi(\sigma))^2, \qquad  \chi(\sigma)=\trace\sigma.\label{diagonal1}
\end{equation}
By the group orthogonality theorem (see e.g., \cite{group})   it follows then  
\begin{equation}
  \D_1=\sum_{\alpha\in \R(\rho)} n^2_\alpha,\label{diagonal2}
\end{equation}
where $n_\alpha$ is the number of times the  irreducible representation  $\alpha$ enters into $\rho$. 
If  $n_\alpha=1$ for each $\alpha$, then $\D_1$ is just  the number of irreducible representations contained in $\rho$.

 \subsection{Non-diagonal contribution}
\begin{figure}[htb]
\begin{center}{

\includegraphics[height=2.0cm]{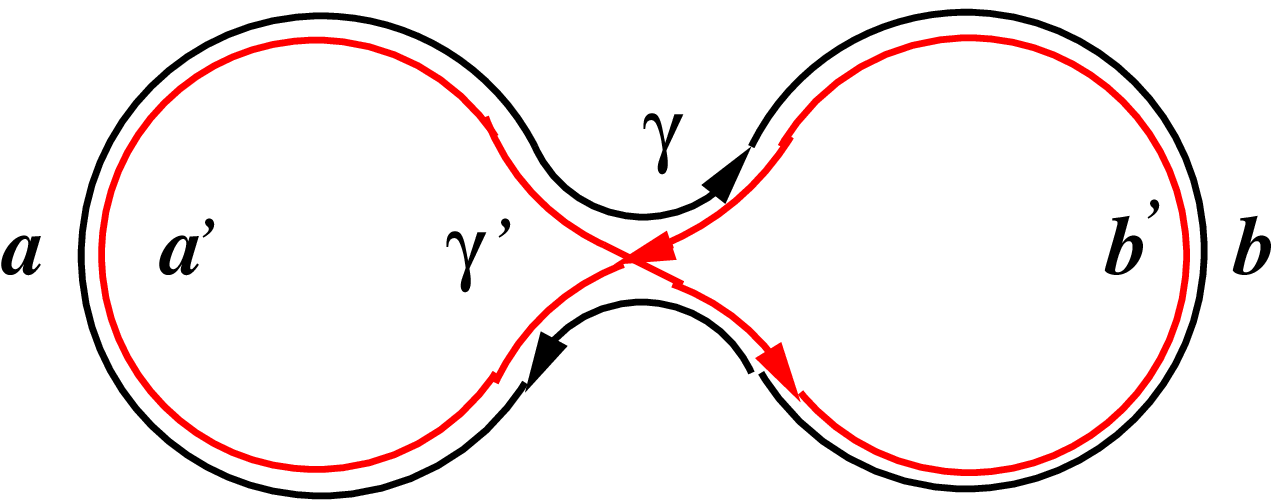}
\hskip 1.0cm
\includegraphics[height=2.0cm]{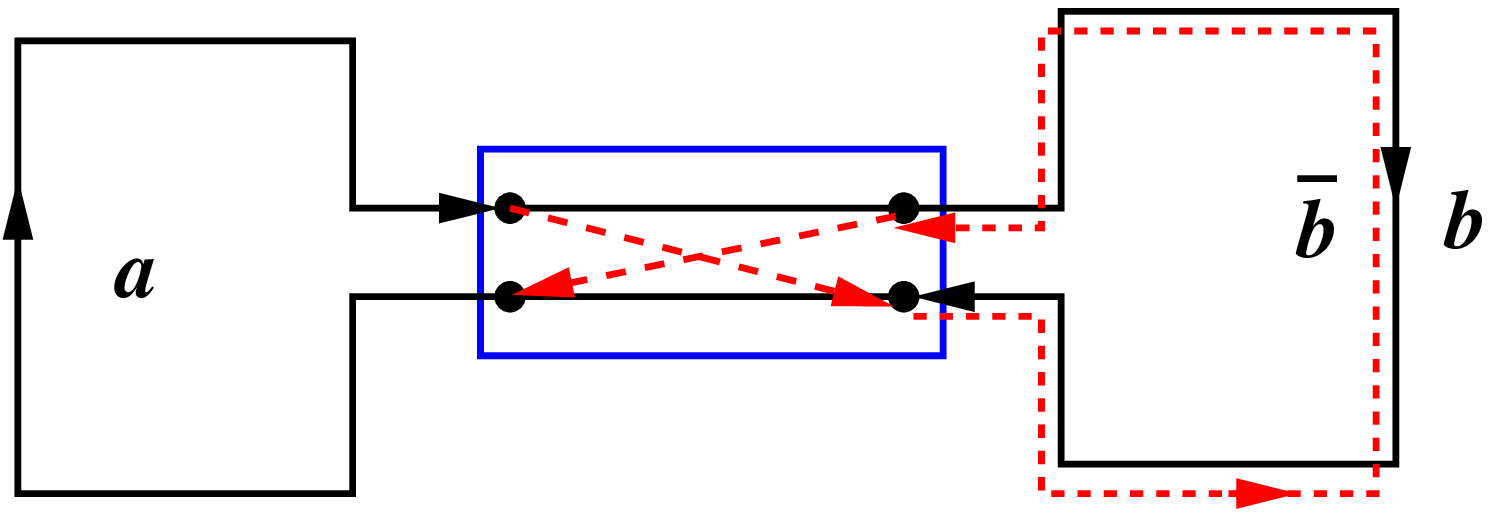}

}\end{center}
\caption{ \small{Sketch of a Sieber-Richter pair in configuration space (left)  and the corresponding diagram (right). 
The two partner periodic orbits $\gamma=ab$, $\gamma'=a\bar{b}$   depicted as solid (black) and dashed (red) lines  follow  each other  at  the stretch ``$a$'', but after leaving the encounter region  (shown as a (blue) rectangle on the right figure)  move in the opposite directions at the stretches ``$b$'' and ``$\bar{b}$'', respectively. } }\label{fig3}
\end{figure}

The second order term in eq.~(\ref{formfac1}) comes from the correlations of periodic orbits  shown in fig.~\ref{fig3}.  These periodic orbits  can be represented as  unions of two (directed) stretches:  $a\cup b$, $a'\cup b'$,
where $a$ is connected with $b$ (resp.   $a'$ with $b'$) 
 at the encounter region. Note that in the configuration space $a$ and   $a'$ are running close to each other. The same holds true  for the stretches $b$ and $ b'$ which have, however,  opposite orientations. Schematically, it is convenient to denote such correlating periodic orbits as 
$\gamma=ab$, $\gamma'=a\bar{b}$, where the ``bar'' symbol stands for an ``opposite orientation''.
Such  structure of  $\gamma$, $\gamma'$  implies that  the pairs of  group elements  $(\sigma_{\gamma},  \sigma_{\gamma'})\in G^{(s)}$  in eq.~(\ref{dnformula}) can be represented in the following form:
\begin{equation}
 \sigma_{\gamma}=gh, \qquad \sigma_{\gamma'}=gh^{-1},
\end{equation}
 where $g$, $h$ correspond to the stretches  $a$  and  $b$, respectively.  It is, therefore, necessary to calculate the following group average:
 \begin{equation}
 \D_2=\frac{1}{|G|^2} \sum_{h\in G}\sum_{g\in G} \chi(gh)\chi(gh^{-1}).
\end{equation}
This quantity can be easily evaluated using the group orthogonality theorem.
\begin{multline}
 \D_2=\frac{1}{|G|^2} \sum_{h,g\in G} \sum_{i,j,k=1}^N\sum_{\alpha,\beta\in \R(\rho)}\rho^{(\alpha)}_{i,j}(g)\rho^{(\alpha)}_{j,i}(h^2)\rho^{(\beta)}_{k,k}(g)\\
=\frac{1}{|G|}\sum_{\alpha\in \R(\rho)}\frac{1}{m_\alpha}\sum_{h\in G}\chi^{(\alpha)}(h^2)=\sum_{\alpha\in\Rr(\rho)}\frac{n^2_{\alpha}}{m_\alpha}-\sum_{\beta\in\Rpr(\rho)}\frac{n^2_{\beta}}{m_\beta},
\end{multline}
where the indices in the last two sums   run over the set  of  real $\Rr(\rho)$ and pseudoreal $\Rpr(\rho)$ irreducible representations entering  $\rho$.  

\begin{figure}[htb]
\begin{center}{
(a)
\includegraphics[width=3.5cm]{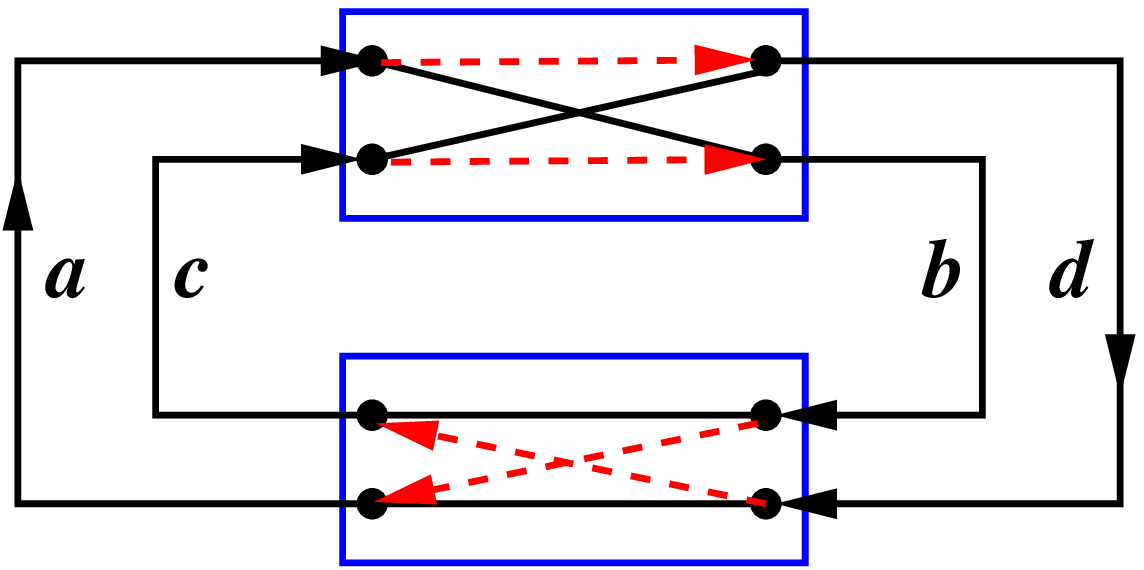}
\hskip 1.0cm
(b)
\includegraphics[width=3.2cm]{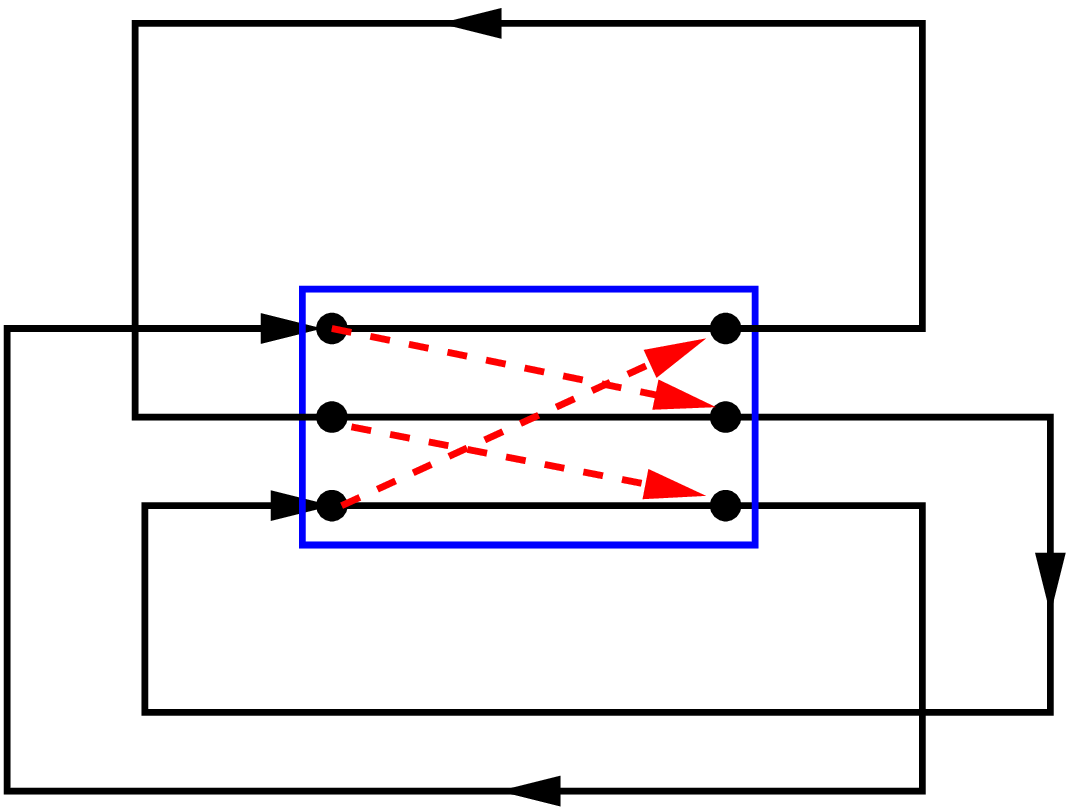}\\
\hskip 0.0cm
(c)
\includegraphics[width=3.5cm]{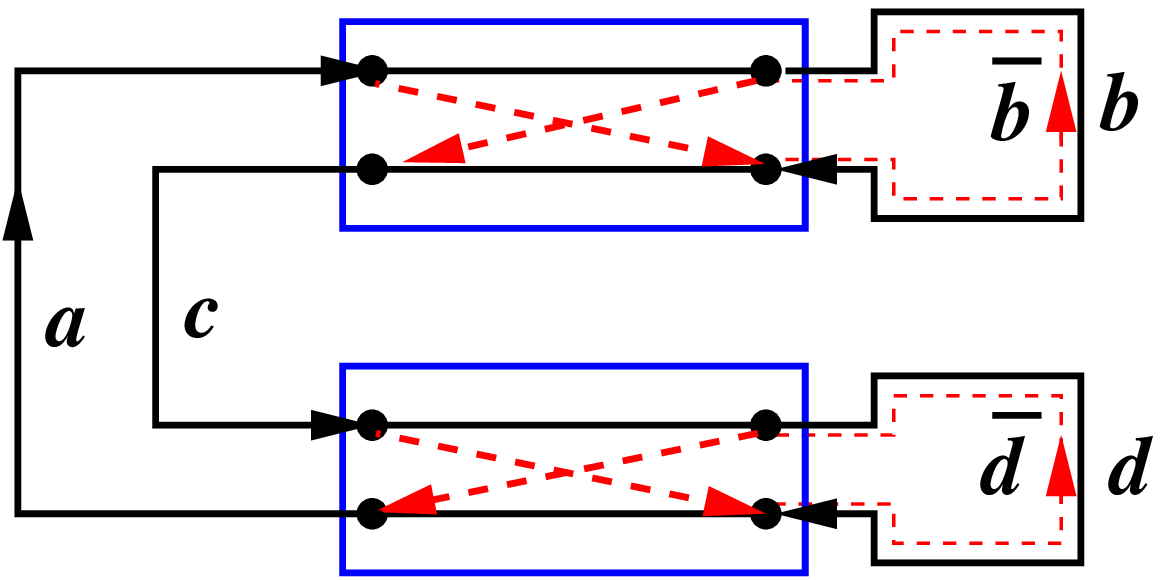}
\hskip 0.5cm
(d)
\includegraphics[width=3.5cm]{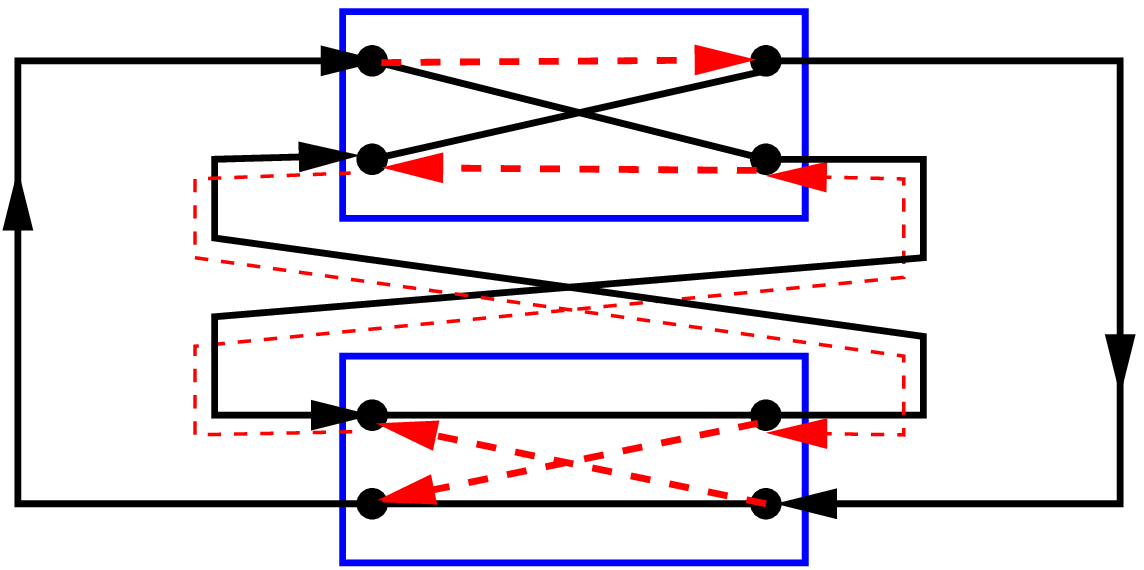}
\hskip 0.5cm
(e)
\includegraphics[width=3.2cm]{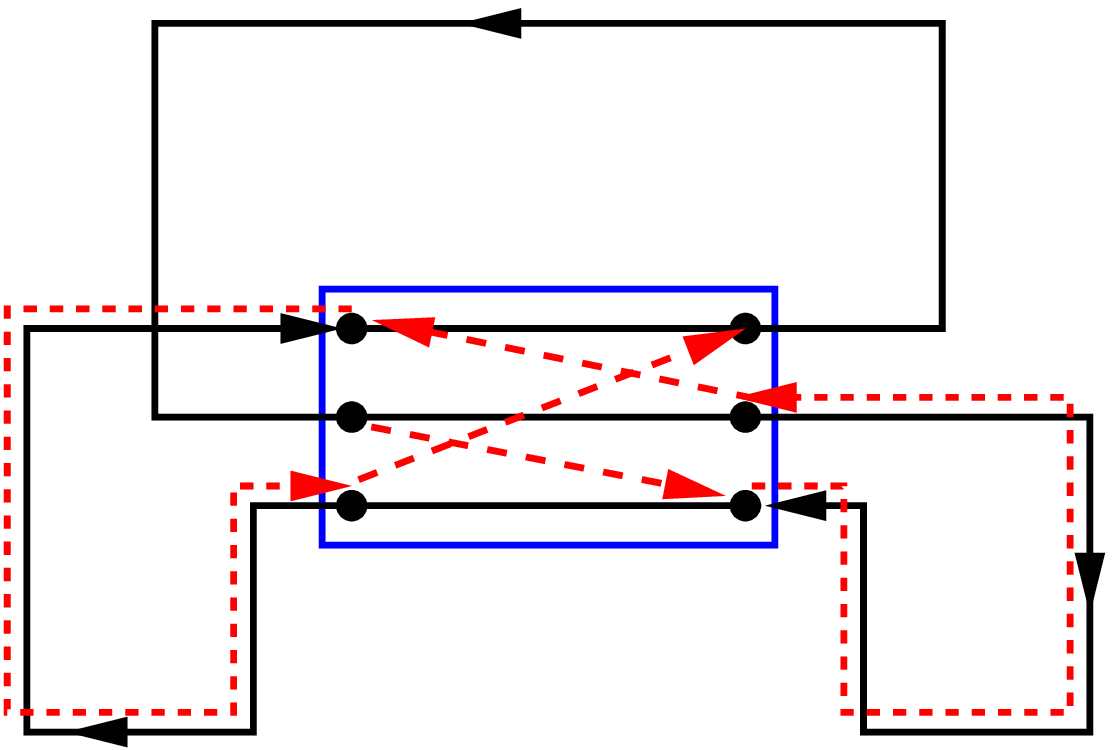}

}\end{center}
\caption{ \small{Five diagrams corresponding to topologically different families of correlating periodic orbits which  contribute to the $\tau^3$ term of the form factor.  The structures depicted at figures (c,d,e) appear only for systems with time reversal invariance.}  }\label{fig4}
\end{figure}

To calculate $\tau^3$ term  of $K(\tau)$ one needs to take in account several different structures of the correlating orbits, which are shown in fig.~\ref{fig4}. In general all such  structures  can be separated into two categories: structures which are relevant both for systems with and without  time reversal invariance and  structures which are relevant only  when time reversal invariance is present. 
The first category is composed of two ``uni-directional'' structures  shown in figs.~\ref{fig4}a,b. Here all correlating     trajectories
have the same direction at each of the encounters. The second  category is represented  by three ``bi-directional'' structures shown in figs. \ref{fig4}c,d,e, where correlating trajectories   have  different directions at least in one of the encounters.
As we show bellow, the expression for $\D_3^{(s)}$  essentially depends on the  type of the  structure $s$.  
 Let us  consider for example  the ``bi-directional'' diagram shown in fig. \ref{fig4}c. In that case the two correlating orbits have the structure  $\gamma=abcd$   and $\bar\gamma=a\bar{d}c\bar{b}$, respectively. As before, we can use the group  orthogonality theorem:
\begin{multline}
 \D^{(c)}_3=\frac{1}{|G|^4} \sum_{g,h\in G}\sum_{f,e\in G}\chi(gfhe)\chi(ge^{-1}hf^{-1})\\
=\frac{1}{|G|^3} \sum_{g,h,f\in G}
 \sum_{\alpha\in \R(\rho)}\frac{n^2_{\alpha}}{m_\alpha}\chi^{(\alpha)}(g^2fh^2f^{-1})\\=\frac{1}{|G|^2}  \sum_{\alpha\in \R(\rho)}\frac{n^2_{\alpha}}{m^2_\alpha}\sum_{g,h\in G}\chi^{(\alpha)}(g^2)\chi^{(\alpha)}(h^2)=\sum_{\alpha\in \Rr(\rho)} \frac{n^2_{\alpha}}{m^2_\alpha}+\sum_{\beta\in\Rpr(\rho)}\frac{n^2_{\beta}}{m^2_\beta}.\label{bi}
\end{multline}
It is straightforward to see that the same result holds for all ``bi-directional'' diagrams of order  $\tau^3$, i.e., $ \D^{(c)}_3=  \D^{(d)}_3=  \D^{(e)}_3$.  On the other hand, for the ``uni-directional`` diagram  in fig.~\ref{fig4}a one has $\gamma=abcd$   and $\bar\gamma=adcb$. This leads to 
\begin{multline}
 \D^{(a)}_3=\frac{1}{|G|^4} \sum_{g,h\in G}\sum_{f,e\in G}\chi(gfhe)\chi(gehf)\\
=\frac{1}{|G|^4} \sum_{g,h\in G}\sum_{f,e\in G}\chi(gfhe)\chi(f^{-1}h^{-1}e^{-1}g^{-1})=\sum_{\alpha\in \R(\rho)} \frac{n^2_{\alpha}}{m^2_\alpha}.\label{uni}
\end{multline}
The same result $\D^{(a)}_3=\D^{(b)}_3$ holds for the diagram on fig.~\ref{fig4}b. Note that both expressions   (\ref{bi},\ref{uni})  have the same form with the notable difference of the range of  irreducible representations $\alpha $  appearing there. Namely,  the sum in   eq.~(\ref{uni})  runs only  over  real and pseudoreal irreducible representations  entering $\rho$ while  the sum  in eq.~(\ref{bi})   includes complex irreducible representations, as well.

Furtheremore, using the same approach  it is straightforward to see that,  for   a general  diagram    of order $\tau^n$   the corresponding  coefficients $\D^{(s)}_n$ are given by
\begin{equation}
 \D^{\mathrm{uni}}_n=\sum_{\alpha\in \R(\rho)} \frac{n^2_{\alpha}}{m^{n-1}_\alpha},\qquad  \D^{\mathrm{bi}}_n=\sum_{\alpha\in \Rr(\rho)} \frac{n^2_{\alpha}}{m^{n-1}_\alpha}+\sum_{\beta\in\Rpr(\rho)}\frac{n^2_{\beta}}{(-m_\beta)^{n-1}},\label{coef}
\end{equation}
for diagrams with ``uni-directional'' and ``bi-directional''  structures, respectively. Substituting  (\ref{coef}) into  eq.~(\ref{formfac1}) and taking into account that  $K_0(\tau)=\KGOE(\tau)$  for a generic $\OmegaO$ with chaotic dynamics, we obtain:
\begin{multline}
 K(\tau)= 
\sum_{\alpha\in \Rr(\rho)} n^2_{\alpha}\left(\frac{m_\alpha }{N}\right)\KGOE\left(\frac{N}{m_\alpha}\tau\right)\\
+\sum_{\beta\in \Rpr(\rho)}(2 n_{\beta})^2 \left(\frac{m_\beta }{2N}\right)\KGSE\left(\frac{2N}{m_\beta}\tau\right)
+\sum_{\nu\in \Ri(\rho)} 2 n^2_{\nu}\tau. \label{finalformfactor}
\end{multline}
Note, that since $\rho_{i,j}(g)$ are real matrices, for every complex representation $\nu$ entering $\rho$  the corresponding complex conjugate   representation $\bar{\nu}$ enters $\rho$, as well. As a result, the last sum in (\ref{finalformfactor}) can be  cast into the form
\begin{equation}
\sum_{(\nu,\bar{\nu})\in \Ri(\rho)} (2 n_{\nu})^2\left(\frac{m_\nu }{N}\right)\KGUE\left(\frac{N}{m_\nu}\tau\right),\label{finalformfactor1}
 \end{equation}
where the sum runs over pairs of all complex representations and their conjugate counterparts.
Eqs.~(\ref{finalformfactor}, \ref{finalformfactor1}) suggest    the following spectral structure  of $\DeltaO$:

\begin{prop} \label{propo}
Let $\rho$ be the standard representation of the structural group $G$ (as defined in Sec.~2) and let
 \[\rho=\bigoplus_{\alpha\in\Rr(\rho)}[\alpha]\bigoplus_{\beta\in\Rpr(\rho)}[\beta]\bigoplus_{\nu\in\Ri(\rho)}[\nu],\]
be its decomposition into a number of real, pseudoreal  and complex  irreducible representations. Then 
  the spectrum of $\DeltaO$  can be  split accordingly:

i) For each real representation $\alpha$  entering $n_\alpha$ times  $\rho$  there exists an associated  GOE-like subspectrum with the density $\bar{d}_{\alpha}=\bar{d}\left(\frac{m_\alpha}{N}\right)$  and the number of degenerate levels $n_\alpha$.

ii) For each pseudoreal representation $\beta$ entering $n_\beta$ times $\rho$ there exists an associated  GSE-like subspectrum  with the dencity $\bar{d}_{\beta}=\bar{d}\left(\frac{m_\beta}{2N}\right)$  and the number of degenerate levels $2n_\beta$.

iii) For each pair of complex  conjugate representations $(\nu,\bar{\nu})$ entering $n_\nu$ times $\rho$ there exists an associated   GUE-like subspectrum with the density of levels  $\bar{d}_{\nu}=\bar{d}\left(\frac{m_\nu}{N}\right)$ and the number of degenerate levels $2n_\nu$.
\end{prop}
In the next section we analyse the origin    of this   spectral decomposition   in cellular billiards. 


\section{Spectral decomposition}
   
Before turning to the  general  case, let us   consider, as an example, the billiards shown in fig.~\ref{fig}. By eq.~(\ref{diagonal1}) the leading order term of the form factor can be straightforwardly evaluated giving  $ \D_1=2$, $ \D_1=1$ for the billiards in fig.~\ref{fig}a and fig.~\ref{fig}b, respectively.  This can be understood, as an indication that the spectrum of the first 
billiard    is composed of two independent GOE components, while   the spectrum of the second billiard   has a single  GOE component.  As we show below, this is indeed so, since  for the billiard in fig.~\ref{fig}a  it is actually possible to find
a projection  operator $P$  commuting with $\DeltaO$. 
To construct such an operator,
consider   a continuous function   $\psi\in C^2(\Omega)$  on $\Omega$ satisfying the same boundary conditions as in (\ref{qbilliard}). Let $\{\psi_1, \psi_2, \psi_3\}$ be the restrictions  of $\psi$ on $\Omega_i, i=1,2, 3$.      Regarding  $(\psi_1, \psi_2, \psi_3)$ as the components of the three-dimensional  vector,  define  the new  set of functions $\{\psi'_1, \psi'_2, \psi'_3\}$  on $\Omega_i, i=1,2, 3$:
\begin{equation}
  \psi'_i=\sum_{j=1}^3 p_{ij}\psi_j, \qquad
p=\frac{1}{3}\begin{pmatrix}
 -1&1&-1\\1&-1&1\\-1&1&-1
\end{pmatrix}.
\end{equation}
We can now lift $\{\psi'_1, \psi'_2, \psi'_3\}$ to  the  new function $\psi'$ on $\Omega$, whose restrictions  on  $\Omega_i, i=1,2, 3$ are  given  by  $\psi'_i$'s.
It is easy to see that $\psi'$ is, in fact, continuous function on $\Omega$ satisfies the same boundary conditions as $\psi$. As a result, the map   $P:\psi\to\psi'$ defines  the linear operator $P$ which acts on the domain $\mathrm{Dom}(\DeltaO)$ of $\DeltaO$.   Since   any solution of eq.~(\ref{qbilliard}) is mapped by $P$ into another solution of this equation  we have $[P, \DeltaO]=0$. Furthermore,   the property $p^2=p$ implies that $P$  is the projection i.e., $P^2=P$.

Turning now to the    general case, for each representation $\alpha\in\R(\rho)$ entering $\rho$  define the following $N\times N$ matrix:
\begin{equation} p^{(\alpha)}=\frac{m_\alpha}{|G|}\sum_{\sigma\in G}\chi^{(\alpha)}(\sigma)\rho^*(\sigma).\label{palpha} 
\end{equation}
By the group orthogonality theorem   these matrices satisfy
  $p^{(\alpha)}p^{(\beta)}=p^{(\alpha)}\delta_{\alpha,\beta}$, for ${\alpha},\beta\in \R(\rho)$ and $\sum_{\alpha\in\R(\rho)} p^{(\alpha)} =1_{N\times N}$. Furthermore, it is straightforward to check that the projections   (\ref{palpha}) 
  commute  with     $\rho(\sigma)$ for all $\sigma\in G$:
\[[p^{(\alpha)},  \rho(\sigma)]=0,  \quad {\alpha}\in \R(\rho).\]

We can now use $p^{(\alpha)}$'s in order  to construct   projection operators $P_\alpha$'s  commuting  with $\DeltaO$.  For a given state $\psi\in\H:=\mathrm{Dom}(\DeltaO)$, with the restrictions $\{\psi_1,\dots  \psi_N\}$ on $\Omega_i$, $i=1,\dots N$ let $\psi'\in\H$ be the state  whose restrictions  on  $\Omega_i$, $i=1,\dots N$ are given by
\begin{equation}
 \psi'_i=\sum_{j=1}^N p^{(\alpha)}_{i,j} \psi_j.
\end{equation}
With each $p^{(\alpha)}$ we  associate  the linear  operation $P_{\alpha}$ which maps  $\psi$ into $\psi'$. It follows  from  the definition of $P_{\alpha}$ and the corresponding  properties of $p^{(\alpha)}$ that $P_{\alpha}P_{\beta}=P_{\alpha}\delta_{\alpha,\beta} $ for any $\alpha,\beta\in\R(\rho)$, $\sum_{\alpha\in\R(\rho)}P_{\alpha}=\mathsf{1}$
and 
\[ [P_{\alpha},\DeltaO]=0, \qquad \alpha\in\R(\rho).\]
Using these projection operators  we can now  split  $\DeltaO$  into the direct sum  
\begin{equation}
 \DeltaO =\bigoplus_{\alpha\in\R(\rho)}\DeltaO^{(\alpha)}, \qquad \DeltaO^{(\alpha)} := P_\alpha \DeltaO P_\alpha, \label{decomposition}
\end{equation}
 where each $\DeltaO^{(\alpha)}$ acts on the subspace $ \H_\alpha=P_\alpha\H$, $\H=\bigoplus_{\alpha\in\R(\rho)} \H_\alpha$.

Now, let us   analyse  degeneracies in the spectrum of each  $\DeltaO^{(\alpha)}$. To this end    note that if $\alpha$ enters $n_\alpha$ times into $\rho$, the projection $p^{(\alpha)}$ can be split further into the sum
 $p^{(\alpha)}=\sum_{i=1}^{n_\alpha}p^{(\alpha)}_i$, such that  the subspaces $h_i^{(\alpha)}:=p_i^{(\alpha)}h$, $h\cong\mathbb{C}^N$,  $i=1,\dots n_\alpha$ are orthogonal to each other and remain invariant under the action of $\rho$. Furtheremore, in this case there exists a group of unitary matrices $u^{(\alpha)}$  commuting with $\rho(g)$ for all $g\in G$ which  mix different subspaces $h^{(\alpha)}_i$'s inside   $h^{(\alpha)}=h^{(\alpha)}_1\oplus \dots \oplus  h^{(\alpha)}_{n_\alpha}$, but leave  every vector $v$ orthogonal to  $ h^{(\alpha)}$ intact: $u^{(\alpha)}v=v$.
   By using  previous arguments, we can lift  the matrices  $p_i^{(\alpha)},u^{(\alpha)}$  to the linear operators $P_{i,\alpha}, U_{\alpha}$ acting  on the Hilbert space  $\H$. From this follows immediately  that the Hilbert space $\H_\alpha$ can be  split into the direct sum $\H_\alpha=\bigoplus^{n_\alpha}_{i=1} \H_{i,\alpha}$, $ \H_{i,\alpha}=P_{i,\alpha}\H$
, where $[P_{i,\alpha},\DeltaO]=0$,  $P_{i,\alpha}P_{j,\alpha}=\delta_{i,j}$, $i=1,\dots n_\alpha$ and there is a group of unitary operators $U_{\alpha}$,  $[U_{\alpha},\DeltaO]=0$ which mix different  $\H_{i,\alpha}$ and leave states from $\H_{\alpha'}$ intact if $\alpha'\neq\alpha$. In its turn this implies that the spectrum of each $\DeltaO^{(\alpha)}$ is at least $n_\alpha$ times degenerate.

\begin{rem}
 For any real representation $\alpha$,  the degeneracy of  the corresponding  spectral component   is given (generically) by the number of times  $\alpha$ enters $\rho$. 
It follows, however, from Proposition~\ref{propo} that for complex and pseudoreal representations there should be additional double degeneracies in the spectrum.  Indeed, for each  pair of  complex conjugate  representations $(\nu, \bar{\nu})$ entering $\rho$, the set of eigenvectors of   $\DeltaO^{(\nu)}$ is  mapped into the   set of orthogonal eigenvectors of   $\DeltaO^{(\bar{\nu})}$ (and vice versa)
by the complex conjugation operation.
Since all eigenvectors of  $\DeltaO$ can be chosen to be real,  $\DeltaO^{(\nu)}$ and  $\DeltaO^{(\bar{\nu})}$ must have the same spectrum. For every pseudoreal representation $\beta$, there exists a unitary operator $A$, such that $A\bar{A}=-1$, where  $\bar{A}$ is the complex conjugate of $A$  and $\beta(g)=A\bar{\beta}(g)A^{-1}$ for any $g\in G$, see \cite{group}. Combining $A$ with the complex conjugation operation $C v=\bar{v}$ we obtain the antiunitary operator $t=AC$ satisfying  $t^2=-1$ and commuting with $\beta(g)$ for all $g\in G$. In its turn, this induces the antiunitary operator $T$ acting on  $\H$, such that  $[\DeltaO^{(\beta)}, T]=0$ and $T^2=-1$. By the last property   vectors  $\psi$ and $T\psi$ must be orthogonal to each other for any $\psi\in\H$,  which implies the double degeneracy  of the spectrum of $\DeltaO^{(\beta)}$ (Kramers' degeneracy), see e.g., \cite{haake}.

\end{rem}


\section{Trace formula for subspectra }

By the decomposition (\ref{decomposition}) the whole spectrum of $\DeltaO$ can be represented as the union of spectra of the operators  $\DeltaO^{(\alpha)}$:
\begin{equation}
\spec(\DeltaO)=\bigcup_{\alpha\in\R(\rho)}\spec(\DeltaO^{(\alpha)}).
\end{equation}
It  is therefore  of interest to obtain a semiclassical expression for the spectral density  of each $\DeltaO^{(\alpha)}$ individually:
\begin{equation}
d^{(\alpha)}(E)=-\frac{1}{\pi}\Im \trace\left(P_\alpha\frac{1}{E+i\varepsilon-\DeltaO}\right).
\end{equation}
To this end we can use  the same approach, as in  the case of systems with geometric symmetries \cite{rob}.  The starting point here is the following representation  of the projected Green's function:
\begin{equation}
 \Gg_{\alpha}(E,x,x):=\left\langle x\left | P_\alpha\frac{1}{E+i\varepsilon-\DeltaO} \right| x\right\rangle=\int_\Omega dy\,\left\langle y |\, P_\alpha \, |x\right\rangle \Gg (E,x,y),\label{greenalpha1}
\end{equation}
where $\Gg(E,x,y)$ stands for the Green's function in the  billiard $\Omega$. Let  $x^{(k)}$ denote the mirror image of the point $x$ in the  domain  $ \Omega^{(k)}$, $k=1,\dots N$  with $x$ being equal to $x^{(m)}\in \Omega^{(m)}$, for some $m$. Using then the definition  (\ref{palpha}) we obtain from eq.~(\ref{greenalpha1})
\begin{multline}
\Gg_{\alpha}(E,x,x)=\sum_{k=1}^N p^{(\alpha)}_{k,m} \Gg (E,x^{(m)},x^{(k)})\\
=\frac{m_\alpha}{|G|}\sum_{\sigma\in G}\chi^{(\alpha)}(\sigma)
\sum_{k=1}^N\rho_{k,m}^*(\sigma)
\Gg (E,x^{(m)},x^{(k)}). \label{greenalpha}
\end{multline}
In order to calculate the oscillating part  $d_{\osc}^{(\alpha)}$ of  the spectral density
\begin{equation}
 d^{(\alpha)}(E)=\bar{d}^{(\alpha)}(E)+d_{\osc}^{(\alpha)}(E)=-\frac{1}{\pi}\Im\int_\Omega dx\, \Gg_{\alpha}(E,x,x),\label{density}
\end{equation}
we can now use the  standard semiclassical  representation for the Green's function $\Gg$ (see e.g., \cite{haake}): 
\begin{equation}
\Gg_{\sem}(E,x^{(m)},x^{(k)})=\frac{1}{i\hbar}\sum_{\tilde{\gamma} [x^{(m)}\to x^{(k)}]}A_{\tilde{\gamma}} \exp{\left(\frac{i}{\hbar}S_{\tilde{\gamma}}(E)\right)},
\end{equation}
where the sum runs over  trajectories in $\Omega$  connecting $x^{(m)}$ to $x^{(k)}$, and $A_{\tilde{\gamma}}$, $S_{\tilde{\gamma}}$ stand for their stability factors and  actions, respectively. Note that the  above expression    can be also rewritten as a sum over closed  trajectories $\gamma[x\to x]$ in the billiard $\OmegaO$:
\begin{equation}
\Gg_{\sem}(E,x^{(m)},x^{(k)}) =\frac{1}{i\hbar}\sum_{\gamma[x\to x]}\rho_{m,k}(\sigma_\gamma)A_{\gamma} \exp{\left(\frac{i}{\hbar}S_{\gamma}(E)\right)}, \label{greensem}
\end{equation}
with $\sigma_\gamma$ being the permutation matrix (\ref{sigmadef}) corresponding to the trajectory $\gamma$.
Substituting now (\ref{greensem}) into (\ref{greenalpha}) and performing saddle point approximation in eq.~(\ref{density}) we obtain for the oscillating part of the spectral density:
\begin{equation}
 d_{\osc}^{(\alpha)}(E)=\frac{m_\alpha}{\pi\hbar|G|}\Re\left\{ \sum_{\sigma}\sum_{\gamma\in\mathrm{PO(\OmegaO)}} \chi^{(\alpha)}(\sigma)
\chi(\sigma^{-1}\sigma_\gamma) \A_{\gamma}\exp{\left(\frac{i}{\hbar}S_{\gamma}(E)\right)} \right\}.
 \end{equation}  
Using the group orthogonality theorem we can  perform   summation over  $\sigma$ and finally get 
\begin{equation}
 d_{\osc}^{(\alpha)}(E)= \frac{n_\alpha}{\pi\hbar} \Re\left\{ \sum_{\gamma\in\mathrm{PO(\OmegaO)}}
 \chi^{(\alpha)}(\sigma_\gamma) \A_{\gamma}\exp{\left(\frac{i}{\hbar}S_{\gamma}(E)\right)} \right\}.
 \end{equation}  

The leading order of the mean spectral density $\bar{d}^{(\alpha)}(E)$ can be also  obtained from eq.~(\ref{greenalpha}) by the integration of the  imaginary part of the Green's function over the points $x^{(m)}= x^{(k)}$:
 \begin{multline}
\bar{d}^{(\alpha)}(E)=-\frac{m_\alpha}{\pi|G|}\sum_{\sigma\in G}\chi^{(\alpha)}(\sigma)
\sum_{k=1}^N\rho_{k,k}^*(\sigma)
\int dx^{(k)}\Im \, \Gg (E,x^{(k)},x^{(k)})\\
=\frac{ m_\alpha \bar{d}}{N |G|}\sum_{\sigma\in G}\chi^{(\alpha)}(\sigma)\chi^{*}(\sigma) +O(E^{-1/2})=n_\alpha \left(\frac{ \bar{d} m_\alpha}{N}\right) +O(E^{-1/2}),
 \end{multline}
where $\bar{d}=\mathrm{Area}(\Omega)/4\pi\hbar^2$ is the leading order (Weyl term) of  the mean spectral density of $\Omega$.

It is  worth mentioning that using semiclassical expression  for  ${d}^{(\alpha)}(E)$  one can straightforwardly establish the type of spectral statistics for each sector $\alpha$. To this end, let us  consider the  form factor for crossover   spectral correlations between  two sectors $\alpha,\beta$:
\[K^{(\alpha,\beta)}(\tau)=\frac{1}{\bar{d}_{\alpha}\bar{d}_{\beta}}\int\left\langle {d}_{\osc}^{(\alpha)}\left(E-\frac{x}{2\bar{d}_{\alpha}}\right){d}_{\osc}^{(\beta)}\left(E+\frac{x}{2\bar{d}_{\beta}}\right)\right\rangle_E e^{-2\pi i\tau x} dx,\]
with $\bar{d}_{\alpha}$ (resp. $\bar{d}_{\beta}$) being  the mean density of the  energy levels (multiplets) in the sector $\alpha$    (resp. $\beta$) given in Proposition~\ref{propo}.
Assuming, as before, that the averaging over  $\chi^{(\alpha)},\chi^{(\beta)}$  can be performed independently of  the averaging  over periodic orbit actions,    the problem of calculation $ K^{(\alpha,\beta)}(\tau)$  reduces   to the evaluation of the   group average:
 \begin{multline}
\frac{1}{4}\left\langle\left(\chi^{(\alpha)}(\sigma)+\chi^{(\alpha)}(\sigma^{-1})\right)
\left(
\chi^{(\beta)*}(\bar\sigma) +\chi^{(\beta)*}(\bar\sigma^{-1})\right)\right \rangle_s\\
= \frac{1}{4|G^{(s)}|} \sum_{(\sigma,\bar\sigma)\in G^{(s)}}\left(\chi^{(\alpha)}(\sigma)+\chi^{(\alpha)*}(\sigma)\right)\left(\chi^{(\beta)}(\bar\sigma)+\chi^{(\beta)*}(\bar\sigma)\right),\label{hru}
\end{multline}
where the sum runs over all  pairs $(\sigma, \bar\sigma) \in G^{(s)}$ having the same  structure  $s\in\S_n$.
By the group orthogonality theorem this average  is  equal to  zero if $\alpha\neq\beta$ implying the absence of correlations between different spectral components. It is therefore sufficient to consider  the case   $\alpha=\beta$. As has been explained in Section~3,  
for real and pseudo-real representations the average (\ref{hru})  is given by 
$(1/m_\alpha)^{n-1}$    and by $(-1/m_\alpha)^{n-1}$, respectively   whenever $s$ is a structure contributing to the $n$-th order of  the form factor. 
Applying then the same arguments as in the derivation of eq.~(\ref{finalformfactor}) yields:
\begin{equation}
K^{(\alpha,\beta)}(\tau)= \begin{cases} 
\delta_{\alpha,\beta} n^2_{\alpha}\KGOE\left(\tau\right) & \text{ if $\alpha$ is real},\\
\delta_{\alpha,\beta}   (2n_{\alpha})^2 \KGSE\left(\tau\right)& \text{ if $\alpha$ is pseudo-real},
  \end{cases}
 \end{equation}
where the additional factor $2$ for  pseudo-real representations accounts for the double degeneracy of the spectrum. 
In the case of  complex representations  the average (\ref{hru}) is given by  $\frac{1}{2}(1/m_\alpha)^{n-1}$ for structures $s$ of the uni-directional type and zero, otherwise. This immediately implies that only the diagonal approximation contributes to the form factor which leads to:
 \begin{equation}
K^{(\alpha,\beta)}(\tau)=\delta_{\alpha,\beta} n^2_{\alpha}\KGUE\left(\tau\right), \qquad \text{ if $\alpha$ is complex}.
\end{equation}
Note finally, that summing up the (rescaled) contributions from all sectors $\alpha$ of the spectrum gives  again  the form factor  (\ref{finalformfactor}).


\section{Conclusion}

To summarise, we have shown that with each cellular billiard $\Omega$ and prescribed boundary conditions on $\partial\Omega$ one can associate certain structure group 
$G$ and its standard representation $\rho(G)$. The  characters of $\rho(G)$ determine  multiplicities of  the periodic trajectories in $\Omega$, as well as the phase factors entering the semiclassical trace formula for  the corresponding quantum billiard problem.
The main result is that the spectrum of the Laplacian $\DeltaO $  can be split in a number of uncorrelated subspectra 
in accordance with the structure of $\rho(G)$. Namely, for each irreducible representation $\alpha$ entering $n_{\alpha}$ times into  $\rho(G)$ there exists an associated $n_{\alpha}$-times degenerate subspectrum $\{E^{(\alpha)} \}$ whose mean level density is proportional to the dimension $m_{\alpha}$ of $\alpha$. 
Furtheremore, for billiards with classically chaotic dynamics the spectral statistics of $\{E^{(\alpha)} \}$ are of the GOE type if $\alpha$ is real,  of the GSE type if $\alpha$ is pseudo-real and of  the GUE type if $\alpha$  is complex.  

It is worth  recalling, that the above spectral structure is reminiscent of the spectral structure for systems with a group of geometrical symmetries $H$, where spectrum can be split in accordance with all irreducible representation of $H$. However, one should   be cautioned to take   this analogy too literally, since in the last case, for instance, the spectral degeneracies are determined by the  
dimensions $m_{\alpha}$ of the irreducible representations rather than by multiplicities  $n_{\alpha}$ (which are not even defined in this case).

It  is  natural  to inquire   about  the  connection between the  geometrical structure of  $\Omega$ and  the  spectral  structure of  the corresponding quantum billiard. For a generic  cellular  billiard with a large number of connections between its cells it can be expected that the matrix group $G$ generated by $N\times N$ matrices $\sigma_i, i=1,\dots l$ is typically  maximum possible matrix group ${G}_{\max}$ for a given  $N$.  For exclusively  Neumann boundary conditions on  $\partial\Omega$, ${G}_{\max}$ is the group of $N\times N$ permutation matrices. 
In this case    $\rho({G}_{\max})$  contains precisely two  irreducible representations, implying that $\spec(-\Delta_\Omega)$ is composed of two independent subspectra.
For the Dirichlet (or mixed) boundary conditions on  $\partial\Omega$,  ${G}_{\max}$ is the group of $N\times N$  matrices having the structure of permutation matrices whose elements $\sigma_{i,j}=\pm 1$  take all possible combinations of  positive and negative signs. 
It is easy to check that in this case $\rho({G}_{\max})$ is an irreducible representation itself  and, therefore, no subspectra appear in $\spec(-\Delta_\Omega)$.   
On the other hand, for some specific structures of   $\Omega$, and boundary conditions on  $\partial\Omega$, a richer spectral structure might, in principle, arise.  
Clearly this happens when a cellular  billiard posses  some geometrical symmetry. In this case  the  standard representation $\rho(G)$  must  contain a non-trivial number of irreducible representations of $G$.
  
One can  wonder, whether it is  possible to have a non-trivial spectral structure of $\Omega$  without having any geometrical symmetry in the system. The answer to this question is positive, as can be seen from an obvious example of billiards with the Neumann  boundary conditions. Any such   billiard $\Omega$ contains as subspectrum the  Neumann spectrum of its basic cell $\Omega^0$. In fact, one can use this property  in order to obtain non-symmetric cellular  billiards $\Omega$ with   an arbitrary number of subspectra. Indeed, it is always possible to construct  cellular  billiards $\Omega'$ with the  Neumann boundary conditions  having some  symmetry   $H$. The spectra of these billiards are  composed of  a  number of independent subspectra corresponding to the irreducible representations of $H$. Taking then any such domain $\Omega'$ as a fundamental cell, one can construct a (larger)  
cellular  billiard $\Omega$ with the  Neumann boundary conditions  which has no geometrical symmetries at all. This billiard, however, will contain  a number of independent subspectra provided by the Neumann spectrum of $\Omega'$.

The simple arguments above demonstrate that, in principle, it is possible to construct non-symmetric cellular  billiards whose spectrum is composed of several components. It would be of interest to investigate what kind of spectral structure might appear in a general case. One interesting question in that regard is   whether there exist cellular  billiards   whose spectrum consists of only  GUE  or GSE components.  

\section*{Acknowledgements}
I would like to thank  C. Joyner, S.\ M\"uller and
 M. Sieber for useful discussions and communicating their  results prior to the publication. 
  The financial  support of   SFB/TR12 of the Deutsche Forschungsgemainschaft is acknowledged.

\bibliographystyle{amsplain}

\begin{thebibliography}{99}
\bibitem{bgs} O. Bohigas, M. J. Giannoni, C. Schmit, {\it Phys. Rev. Lett. } {\bf 52}
(1) (1984)

\bibitem{br1}   M. Berry, M. Robnik,  False time-reversal violation and energy level statistics: the role of anti-unitary symmetry, {\it J. Phys. A } {\bf 19}, 669-682 (1986)

\bibitem{lss} F. Leyvraz,  C. Schmit, T. H. Seligman, {\it J. Phys. A:
 Math. Gen.} {\bf 29}, L575-L580  (1996)


\bibitem{rob}  J. M. Robbins, Discrete symmetries in periodic-orbit theory, {\it  Phys. Rev. A:
 } {\bf 40}, 2128-2136  (1989)
\bibitem{lau} B. Lauritzen, Discrete symmetries and periodic-orbit expansion, {\it  Phys. Rev. A} {\bf 43}, 603-606  (1991)
\bibitem{kr} J. P. Keating, J. M. Robbins, Discrete symmetries and spectral statistics, {\it J. Phys. A:
 Math. Gen.} {\bf 30}, L177-L181  (1997)

\bibitem{ms}  C. Joyner, S.\ M\"uller, M. Sieber,  preprint (2010)

\bibitem{bog1}   E. B. Bogomolny, B. Georgeot, M.J. Giannoni,  C. Schmit,
{\it Phys. Rev. Lett.}  {\bf 69}, 1477-1480 (1992)
\bibitem{bog2} E. B. Bogomolny, B. Georgeot, M.J. Giannoni, C. Schmit,
Arithmetical chaos, {\it Phys. Rep.} {\bf 291}, 219-324 (1997)
\bibitem{cat1} J.H.  Hannay, M.V. Berry,  Quantisation of linear maps on the torus -- Fresnel diffraction by a periodic grating {\it Physica D } {\bf 1}, 267-290 (1980)
\bibitem{cat2} J.P. Keating, F. Mezzadri, Pseudo-symmetries of Anosov maps and spectral statistics, {\it Nonlinearity} {\bf 13}, 747-775  (2000)

\bibitem{robnik}  G. Veble, T. Prosen, M. Robnik, {\it New J. Phys.}, {\bf 9}, 15 (2007)
\bibitem{moj} B. Gutkin, {\it J. Phys. A: Math. Theor.} {\bf 40},  F761-F769 (2007)

\bibitem{berry} M. Berry, Semiclassical theory of spectral rigidity {\it Proc. R. Soc. A} {\bf 400}, 229-251 (1985)

\bibitem{kac} M. Kac, {\it Am. Math. Monthly}, {\bf 1} 23 (1966).
\bibitem{gordon} C. Gordon et al, Bull. {\it Am. Math. Soc.,} {\bf 27} 134 (1992).
\bibitem{mojuzy} B. Gutkin, U. Smilansky,  Can one hear the shape of a graph?
{\it J. Phys. A: Math. Gen.} {\bf 34}, 6061-6068 (2001)
\bibitem{rami} O. Parzanchevski,  R. Band, {\it  J. Geom. Anal.} {\bf 20}, 439-471  (2010) 
\bibitem{haake} F. Haake, Quantum Signatures of Chaos, 2nd ed. (Springer-Verlag, Berlin, 2001)
\bibitem{rs} M. Sieber, K. Richter {\it  Phys. Scripta.} {\bf T90}, 128  (2001)
\bibitem{haake1}
S.\ M\"uller, S.\ Heusler, P.\ Braun, F.\ Haake, and A.\ Altland, Phys.\ Rev.\ Lett.\ {\bf93}, 014103 (2004); Phys.\ Rev.\ E {\bf72}, 046207 (2005).

\bibitem{group}
M. Hamermesh, Group Theory and its Applications to Physical Problems, Addison-Wesley, Reading (1962). (Reprinted by Dover).




\end{thebibliography}

\end{document}